\documentclass[11pt]{article}
\usepackage{amssymb}
\usepackage{graphics}
\usepackage{epsfig}
\usepackage{a4wide}

\begin{document}
\title{Full one-loop supersymmetric electroweak corrections to $t\bar{t}h^{0}$
associated production in $e^+e^-$ annihilation\footnote{Supported
by National Natural Science Foundation of China.}} \vspace{3mm}

\author{{ Liu Jing-Jing$^{2}$, Ma Wen-Gan$^{1,2}$, Han Liang$^{2}$, Zhang Ren-You$^{2}$, Jiang Yi$^{2}$ and Wu Peng$^{2}$}\\
{\small $^{1}$ CCAST (World Laboratory), P.O.Box 8730, Beijing 100080, People¡¯s Republic of China} \\
{\small $^{2}$ Department of Modern Physics, University of Science and Technology}\\
{\small of China (USTC), Hefei, Anhui 230027, People¡¯s Republic of China} }

\date{}
\maketitle \vskip 12mm

\begin{abstract}
We present a precise calculation of the lightest neutral Higgs
boson production associated with top-quark pair at a linear
collider. The full one-loop electroweak ${\cal O}(\alpha_{ew})$
contributions to the process $e^+e^- \to t\bar t h^0$ within the
minimal supersymmetric standard model (MSSM) are included. We
analyze the dependence of the electroweak corrections on the MSSM
parameters such as $M_{A^0}$, $\tan\beta$, $M_2$, $A_{f}$,
$M_{SUSY}$ and $\mu$. The results show that the full one-loop
electroweak radiative corrections turn out to be about $-20\%$
quantitatively and thus are important for future $e^+e^-$ linear
colliders.

\end{abstract}


\vfill \eject

\baselineskip=0.36in

\newcommand{\nb}{\nonumber}

\section{Introduction}
\par
To search for Higgs boson is one of the most important tasks of
the experimental programs at future high-energy colliders. As we
know that in the frameworks of the standard model(SM) and its
extensions, electroweak symmetry breaking and mass generation of
gauge bosons and fermions are induced by the Higgs
mechanism\cite{sm1,sm2}. By adopting two Higgs doublets to
preserve the supersymmetry in the minimal supersymmetric standard
model(MSSM), five Higgs bosons($h^0,H^0,A^0,H^\pm$) are predicted.
However, none of the Higgs bosons has been directly explored
experimentally until now, except that LEP2 experiments provided a
lower bound of $114.4~{\rm GeV}$ \cite{ALEPH1} and a upper bound
of $260~{\rm GeV}$ \cite{LEP} for the SM Higgs boson mass at the
$95\%$ confidence level. In representative scans of the parameters
of the MSSM, the mass limit of $m_{h^{0}}>91.0$ GeV is obtained
for the lightest CP-even Higgs boson \cite{ALEPH2}.

\par
The present precise experimental data have shown an excellent
agreement with the predictions of the SM except the Higgs
sector\cite{higgs}. These data strongly constrain the couplings
between gauge boson and fermions, such as ($\lambda_{Zf\bar{f}}$
and $\lambda_{Wf\bar{f'}}$), and the gauge self-couplings, but say
little about the couplings between the Higgs boson and fermions
($\lambda_{Hf\bar{f}}$). In both theories of the SM and the MSSM
the Higgs mechanism predicts the Yukawa coupling, i.e., the
coupling between the Higgs boson and fermions, e.g.,
$\lambda_{h^0f\bar{f}}$, its coupling strength is proportional to
the mass of fermion, except the coupling $\lambda_{h^0f\bar{f}}$
in the MSSM is modified by the mixing angles $\alpha$ and $\beta$.
Because of the heavy top-quark mass, the coupling
$\lambda_{h^0t\bar{t}}$ is the strongest one among all the
Higgs-fermion-antifermion couplings, and the cross section of the
$t\bar{t}h^0$ associated production is dominated by the amplitudes
describing Higgs boson radiation off the top or the anti-top-
quark. Therefore, the process of $t\bar{t}h^{0}$ associated
production at future colliders is not only particularly suitable
in discovering the Higgs boson with the intermediate mass, but
also helpful in measuring the Yukawa coupling strength. However,
to determinate the profile of the Yukawa coupling concretely with
clearer background, an $e^+e^-$ linear collider is necessary. In
fact, there are several linear colliders which have been proposed
and designed, such as TESLA\cite{TESLA}, NLC\cite{NLC},
GLC\cite{JLC} and CERN CLIC\cite{CLIC}. Base on the experimental
precision at the present technique level, the theoretical QCD and
electroweak radiative corrections should be taken into account.
People believe that the precise test for the Higgs sector can be
implemented by means of the future high-energy colliders, such as
the CERN large hadron collider (LHC) and linear colliders (LC's).

\par
Recently, a lot of effort has been invested in improving the
precision of the QCD corrections to the process $p\bar{p}/pp\to
t\bar{t}h^0+X$ theoretically \cite{Reina1, Reina2, Been,
Rainwater}. Considerable progress has been achieved in the
calculations of the electroweak corrections and QCD corrections in
the SM \cite{You,Denner} and MSSM \cite{Wu,Zhu,Spira} to the
process $e^+e^-\to t\bar{t}h^0$. The precise calculations in the
SM for the process $\gamma\gamma \to t\bar{t}h^0$ at the tree
level and the corrections of NLO QCD and one-loop electroweak
interactions, have been presented in Refs. \cite{Cheung,Chen}. The
calculation in Ref.\cite{Wu} has been done by taking into account
the supersymmetric electroweak corrections of the order ${\cal
O}(\alpha_{ew}m_{t,b}^2/m_W^2)$ and ${\cal
O}(\alpha_{ew}m_{t,b}^3/m_W^3)$. In this work we present in detail
the calculation of the full ${\cal O}(\alpha_{ew})$ electroweak
radiative corrections to the process $e^+ e^- \to t\bar{t}h^0$ in
the framework of the MSSM.

\par
This paper is organized as follows: In Sect.2, we present the
calculation of the complete one-loop electroweak radiative
corrections to $e^+e^- \to t\bar{t}h^0$ process in the MSSM. The
numerical results and discussion are given in Sect.3. Finally, we
give a short summary.

\vskip 10mm
\section{Calculation}

In our calculation, we adopt the 't Hooft-Feynman gauge. In the
calculation of loop diagrams we take the definitions of one-loop
integral functions in Ref.\cite{loop}. The Feynman diagrams and
relevant amplitudes are created by $FeynArts~3$ \cite{Feynarts}
automatically, and the Feynman amplitudes are subsequently reduced
by $FORM$ \cite{Form}. Our renormalization procedure is
implemented in these packages. The numerical calculation of
integral functions are implemented by using our Fortran programs,
in which the 5-point loop integrals are evaluated by using the
approach presented in Ref.\cite{pentagon}

\par
Because of the fact that the Yukawa coupling of Higgs/Goldstone to
fermions is proportional to the fermion mass, we ignore the
contributions of the Feynman diagrams which involve the Yukawa
couplings between any Higgs/Goldstone boson and electrons. There
are seven Feynman diagrams relevant to the process $e^+ e^- \to
t\bar{t}h^0$ at the tree level, which are depicted in Fig.1. The
diagrams shown in Fig.1 can be divided into two groups. One
contains the diagrams with Higgs boson strahlung from top or
anti-top-quark final state and the $t-\bar t-h^0$ Yukawa coupling
is thus involved. Another group involves the diagrams with a Higgs
boson radiated via Higgs-gauge boson interactions, such as
$Z^0-G^0-h^0$, $Z^0-A^0-h^0$ and $Z^0-Z^0-h^0$ vertexes, but is
free from the $t-\bar t-h^0$ Yukawa coupling. The electroweak
one-loop Feynman diagrams can be classified into self-energy,
triangle, box and pentagon diagrams. Some of the pentagon diagrams
are depicted in Fig.2 as a representation, in which five point
tensor integrals of rank 4 are involved in the corresponding
amplitudes.

\begin{figure}[htb]
\centering
\includegraphics[scale=0.7]{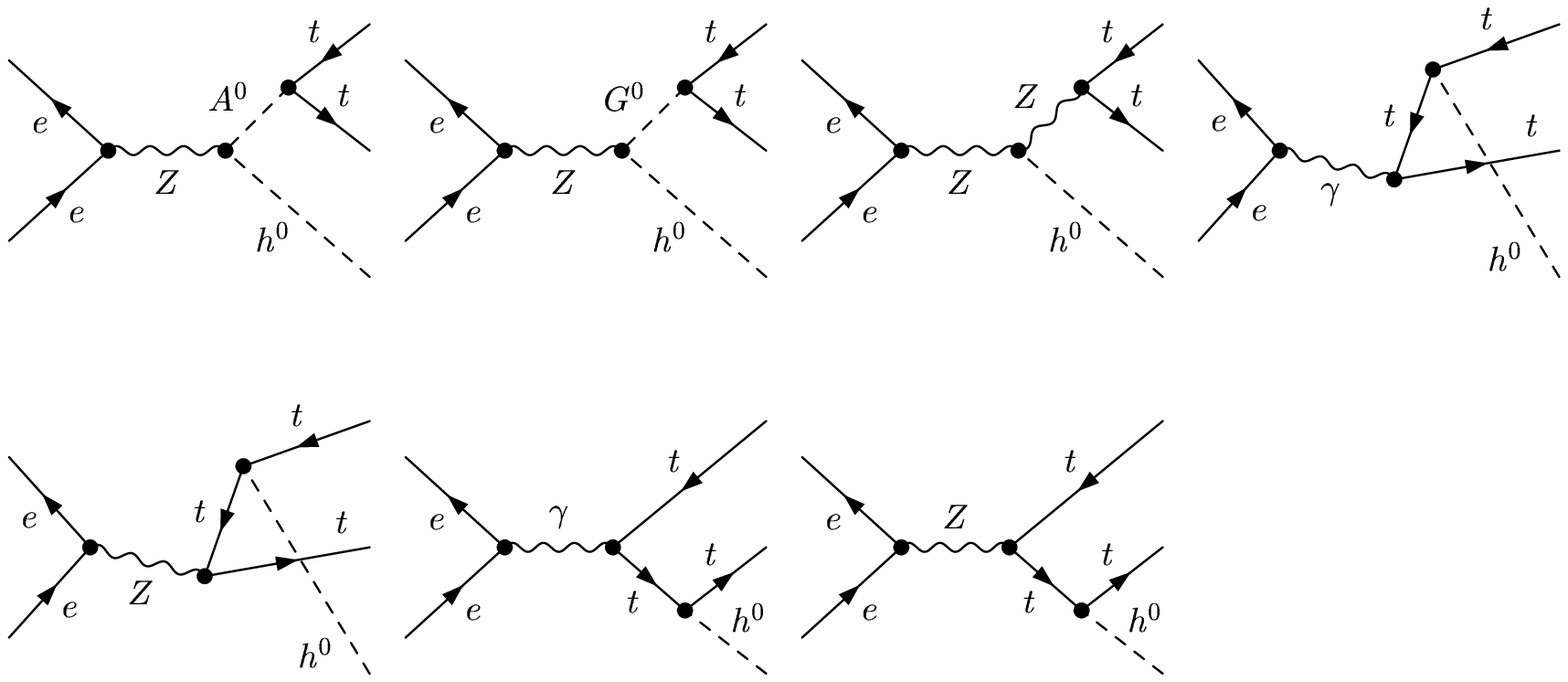}
\caption{The tree level Feynman diagrams of the process $e^+ e^-
\to t\bar{t}h^0$.}
\end{figure}

\begin{figure}[htb]
\centering
\includegraphics[scale=0.75]{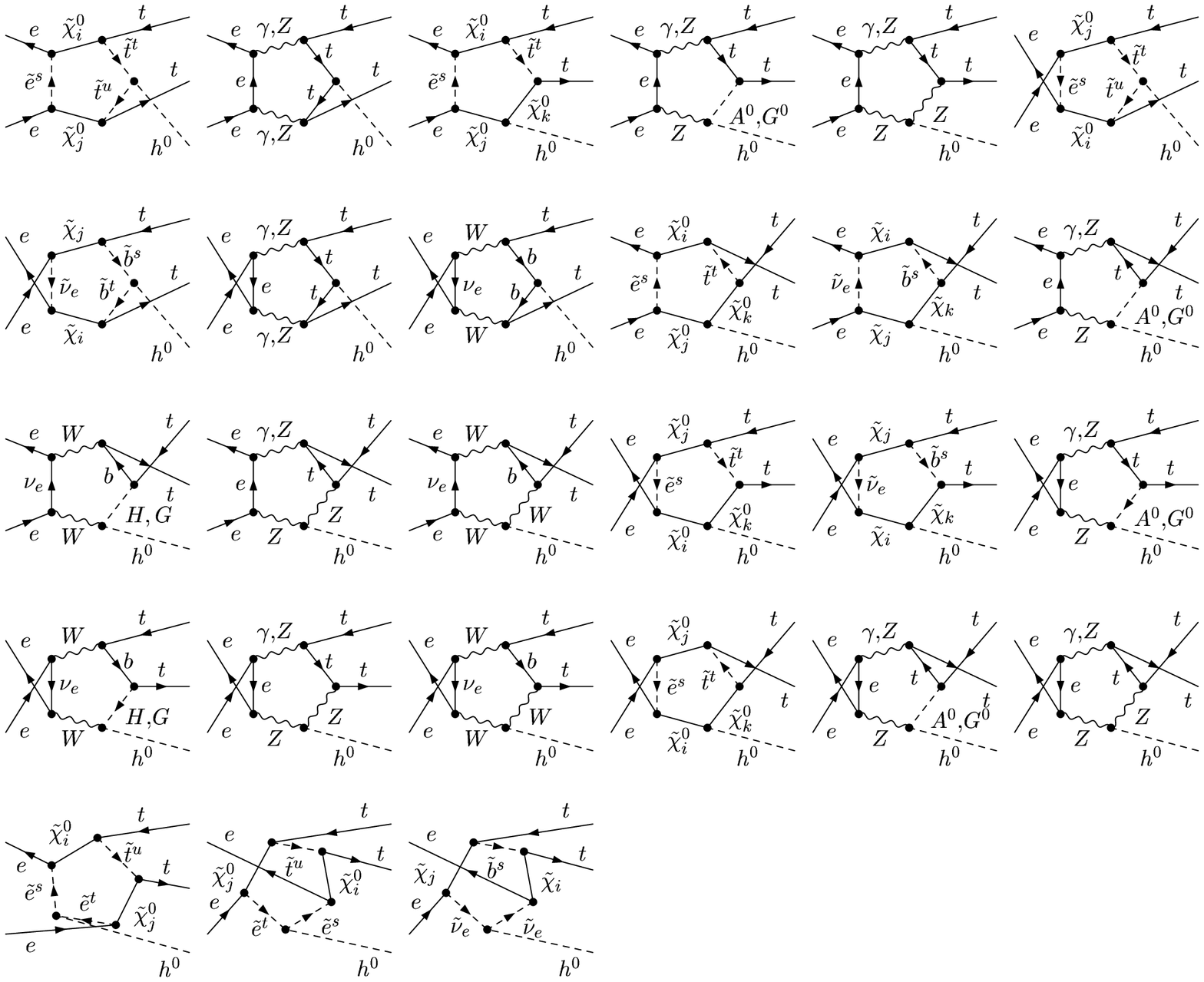}
\caption{Some of the pentagon diagrams of the process $e^+ e^- \to
t\bar{t}h^0$, where $i$, $j$, $k$ are indexes of
neutralino/chargino and $s$, $t$, $u$ are indexes of squarks}
\end{figure}

\par
The ${\cal O}(\alpha_{ew})$ virtual correction to the cross
section for the process $e^+(p_1)+e^-(p_2) \to
t(k_1)+\bar{t}(k_2)+h^0(k_3)$ can be expressed as
\begin{eqnarray}
\sigma_{virtual}=\sigma_{tree}\delta_{virtual}= \frac{(2 \pi )^4
N_c}{2|\vec{p}_1|\sqrt{s}}\int d\Phi_3\overline{\sum_{spin}}
Re({\cal M}_{tree}{\cal M}^{\dag}_{virtual})
\end{eqnarray}
where $N_c=3$ and $\vec{p}_1$ is the c.m.s. momentum of the
initial positron, $d\Phi_3$ is the three-body phase space element,
and the bar over summation recalls averaging over initial spins
\cite{RevD}. $\sigma_{tree}$ and ${\cal M}_{tree}$ are the cross
section and amplitude at the tree level for process $e^+e^-\to
t\bar{t}h^0$ separately. ${\cal M}_{virtual}$ is the renormalized
amplitude from all the electroweak one-loop Feynman diagrams and
the corresponding counterterms.

\par
As we know, the contributions of the virtual one-loop diagrams
contain both ultraviolet (UV) and infrared (IR) divergences. In
this paper, we adopt the dimensional reduction ($DR$)
regularization scheme to preserve supersymmetry, and use the
on-mass-shell conditions (neglecting the finite widths of the
particle) to renormalize fields \cite{ren}. The electric charge of
electron $e$, the physical masses $m_W$, $m_Z$, $M_{A^0}$, $m_t$,
Higgs mixing angle $\alpha$ and the ratio of the vacuum
expectation values $\tan \beta$ are chosen to be the relevant
renormalized parameters. The definitions and the explicit
expressions of these renormalization constants can be found in
Refs. \cite{MSSM}. Here we list them as follow:
\begin{eqnarray}
&&m_{t,0}~=~m_t+\delta
m_{t},~~~~~~~~~~~~~~W^{\pm}_0~=~(1+\frac{1}{2}\delta
Z_W)W^{\pm},  \nonumber \\
& & t_0^{L}~=~(1+\frac{1}{2}\delta Z_{t}^L)t^L, ~~~~~~~~~~~~
t_0^{R}~=~(1+\frac{1}{2}\delta Z_{t}^R)t^R,  \nb \\
& &\left(
\begin{array}{c}
Z^{0({0})} \\ A^{0({0})}
\end{array}
\right)~=~\left(
\begin{array}{cc}
1 + \frac{1}{2} \delta Z_{ZZ} & \frac{1}{2} \delta Z_{ZA} \\
\frac{1}{2} \delta Z_{AZ} & 1 + \frac{1}{2} \delta Z_{AA}
\end{array}
\right) \left(
\begin{array}{c}
Z^0 \\ A^0
\end{array}
\right), \nonumber \\
& &m_{W,0}^2=m_{W}^2+\delta m_W^2,~~~~~~~~~~~~~~m_{Z,0}^2~=~m_{Z}^2+\delta m_Z^2, \nonumber \\
& &M_{A^0}^{({0})2}~=~ M_{A^0}^2 + \delta M_{A^0}^2,~~~~~~~~~~
\alpha^{({{0}})}~=~\alpha + \delta \alpha, \nonumber \\
& &\beta^{({{0}})}~=~\beta + \delta \beta, ~~~~~~~~~~~~~~~~~~~~~
e^{({{0}})} = Z_e e = ( 1 + \delta Z_e ) e,\nb\\
& &\left(
\begin{array}{c}
H^{0({0})} \\ h^{0({0})}
\end{array}
\right)~=~\left(
\begin{array}{cc}
1 + \frac{1}{2} \delta Z_{H^0H^0} & \frac{1}{2} \delta Z_{H^0h^0} \\
\frac{1}{2} \delta Z_{h^0H^0} & 1 + \frac{1}{2} \delta Z_{h^0h^0}
\end{array}
\right) \left(
\begin{array}{c}
H^0 \\ h^0
\end{array}
\right), \nonumber\\
& &\left(
\begin{array}{c}
A^{0({0})} \\ G^{0({0})}
\end{array}
\right)~=~\left(
\begin{array}{cc}
1 + \frac{1}{2} \delta Z_{A^0A^0} & \frac{1}{2} \delta Z_{A^0G^0} \\
\frac{1}{2} \delta Z_{G^0A^0}  & 1 + \frac{1}{2} \delta Z_{G^0G^0}
\end{array}
\right) \left(
\begin{array}{c}
A^0 \\ G^0
\end{array}
\right),\nonumber\\
\delta m_W^2 &=& \widetilde{Re}
\Sigma_T^W(m_W^2),~~~~~~~~~~~~~~~~~~~~~
\delta m_Z^2~=~Re \Sigma_T^{ZZ}(m_Z^2),     \nb\\
\delta Z_W &=& -\widetilde{Re}\frac{\partial
\Sigma_T^W(k^2)}{\partial k^2}|_{k^2=m_W^2},~~~~~~~~~ \delta
Z_{ZZ}~=~-Re\frac{\partial
\Sigma_T^{ZZ}(k^2)}{\partial k^2}|_{k^2=m_Z^2},  \nb\\
\delta M_{A^0}^2 &=&
\widetilde{Re}\Sigma^{A^0A^0}(M_{A^0}^2)-b_{AA}, ~~~~~~~~\delta
Z_{H^0 H^0}~=~-\widetilde{Re} \frac{\partial \Sigma^{H^0
H^0}(k^2)}{\partial k^2} \mid_{k^2 =
m_{H^0}^2}, \nb\\
\delta Z_{h^0H^0}&=&\frac{2}{m_{H^0}^2 - m_{h^0}^2}\widetilde{Re}
[ b_{Hh} - \Sigma^{H^0 h^0}(m_{H^0}^2)], ~\delta Z_{h^0
h^0}~=~-\widetilde{Re} \frac{\partial \Sigma^{h^0
h^0}(k^2)}{\partial k^2} \mid_{k^2 = m_{h^0}^2},\nb \\
\delta Z_{H^0h^0}&=&\frac{2}{m_{h^0}^2 - m_{H^0}^2} \widetilde{Re}
[ b_{Hh} - \Sigma^{H^0 h^0}(m_{h^0}^2)], ~\delta Z_{A^0A^0}~=~
-\widetilde{Re} \frac{\partial \Sigma^{A^0
A^0}(k^2)}{\partial k^2} \mid_{k^2 = M_{A^0}^2},\nb \\
\delta Z_{G^0 G^0}&=&-\widetilde{Re} \frac{\partial \Sigma^{G^0
G^0}(k^2)}{\partial k^2} \mid_{k^2 =0}, ~~~~~~~~\delta Z_{G^0
A^0}~=~\frac{2}{M_{A^0}^2} \widetilde{Re} [ b_{GA} -
\Sigma^{G^0 A^0}(M_{A^0}^2)], \nb\\
\delta Z_{A^0G^0}&=&-\frac{2}{M_{A^0}^2} \widetilde{Re} [ b_{GA} -
\Sigma^{G^0 A^0}(0)].
\end{eqnarray}
The Higgs tadpole parameters $ b_{AA}$, $b_{GA}$, $b_{Hh}$ are
defined and expressed as in Ref. \cite{MSSM}. The operator
$\widetilde{Re}$ takes only the real part of the loop integrals
and does not affect the possible complex couplings. The
renormalization counterterm of nonindependent parameter Higgs
mixing angle $\alpha$ can be obtained by satisfying the tree-level
relation
\begin{eqnarray}
\tan{2\alpha}=\frac{M_{A^0}^2+m_Z^2}{M_{A^0}^2-m_Z^2}\tan{2\beta},
~~~~~~~~-\pi/2<\alpha<0,
\end{eqnarray}
and has the expression as
\begin{eqnarray}
\delta \alpha = \sin 4\alpha \left[\frac{\delta\beta}{\sin 4
\beta}+\frac{M_{A^0}^2 \delta m_Z^2-m_Z^2 \delta M_{A^0}^2}{2
(M_{A^0}^4-m_Z^4)}\right].
\end{eqnarray}
By imposing $\delta v_L/v_L=\delta v_R/v_R$, we get the expression
for the renormalization counterterm of the angle $\beta$ as
\begin{eqnarray}
\delta \beta = \frac{\delta Z_{G^0 A^0}}{4}.
\end{eqnarray}
As we except, the UV divergence contributed by the one-loop
diagrams should be cancelled by the counterterms exactly. Then we
get a UV finite cross section including ${\cal O}(\alpha_{ew})$
virtual radiative corrections. We have verified the cancellation
of the UV both analytically and numerically in our calculation.

\par
The IR divergence in the process $e^+e^-\to t\bar{t}h^0$ comes
from the virtual photonic corrections. It can be exactly cancelled
by including the real photonic bremsstrahlung corrections to this
process in the soft photon limit. The real photon emission process
is denoted as
\begin{equation}
\label{process}e^+(p_1)+e^-(p_2)\to
t(k_1)+\bar{t}(k_2)+h^0(k_3)+\gamma(k),
\end{equation}
where the real photon radiated from the initial electron/positron
and the final top/anti-top-quark, can have either soft or
collinear nature. The collinear singularity is regularized by
keeping nonzero electron mass. $m_\gamma$ is introduced to refer
to a mass regulator for the photonic IR divergencies. In order to
isolate the soft photon emission singularity in the real photon
emission process, we use the general phase-space-slicing method
\cite{Harris}. The bremsstrahlung phase space is divided into
singular and nonsingular regions, and the cross section of the
real photon emission process (\ref{process}) is decomposed into
soft and hard terms
\begin{equation}
\sigma_{real}=\sigma_{soft}+\sigma_{hard}=\sigma_{tree}
(\delta_{soft}+\delta_{hard}).
\end{equation}
where both $\sigma_{soft}$ and $\sigma_{hard}$ depend on the
arbitrary soft cutoff $\Delta E/E_b$, $E_b=\sqrt{s}/2$ is the
electron beam energy in the c.m.s. frame. The total real cross
section $\sigma_{real}$ is independent of the cutoff. Since in our
practical calculation of the $\sigma_{soft}$, the soft cutoff
$\Delta E/E_b$ is set to be very small, the terms of order $\Delta
E/E_b$ can be neglected and the soft contribution can be evaluated
by using the soft photon approximation analytically \cite{Hooft}
\begin{eqnarray}
d\sigma_{soft}=-d\sigma_{tree}\frac{\alpha_{ew}}{2
\pi^2}\int_{|\vec{k}| \leq \Delta E}\frac{d^3k}{2k_0}
\left[\frac{p_1}{p_1\cdot k}-\frac{p_2}{p_2\cdot
k}-\frac{e_tk_1}{k_1\cdot k}+\frac{e_tk_2}{k_2\cdot k}\right],
\end{eqnarray}
where $\Delta E$ is the energy cutoff of the soft photon and
$k_0\leq \Delta E \ll \sqrt{s}$, $e_t=2/3$ is the electric charge
of the top-quark, $k_0=\sqrt{|\vec{k}|^2+m_{\gamma}^2}$ is the
energy of the photon, and $p_1$ and $p_2$ are the four momenta of
$e^+$ and $e^-$ respectively. The IR divergence from the soft
contribution cancels exactly that from the virtual corrections.
Therefore, the sum of the virtual and soft cross sections is
independent of the infinitesimal photon mass $m_\gamma$. The hard
photon emission cross section $\sigma_{hard}$ is UV and IR finite
with the radiated photon energy being larger than $\Delta E$. In
this work, The phase space integration of the process $e^+e^- \to
t\bar t h^0\gamma$ with hard photon emission is performed by using
the program $GRACE$ \cite{grace}. Finally, the total cross section
including the full one-loop electroweak corrections for the
process $e^+e^-\to t\bar{t}h^0$, can be obtained by
\begin{equation}
\sigma_{total}=\sigma_{tree}+\sigma_{virtual}+\sigma_{real}=\sigma_{tree}(1+\delta_{total})
\end{equation}
where $\delta_{total} = \delta_{virtual} + \delta_{soft} +
\delta_{hard}$ is defined as the full ${\cal O}(\alpha_{ew})$
electroweak relative correction.

\vskip 10mm
\section{Numerical results and discussions }
In the numerical calculation, we use the following SM parameters
\cite{RevD}
\begin{eqnarray}
m_e&=&0.510998902~{\rm MeV},~m_\mu~=~105.658369~{\rm MeV},~m_\tau~=~1776.99~{\rm MeV},\nb\\
m_u&=&66~{\rm MeV},~~~~~~~~~~~~~~m_c~=~1.2~{\rm GeV},~~~~~~~~~~~~~m_t~=~178.1~{\rm GeV},\nb\\
m_d&=&66~{\rm MeV},~~~~~~~~~~~~~~m_s~=~150~{\rm MeV},~~~~~~~~~~~~m_b~=~4.3~{\rm GeV} ,\nb\\
m_W&=&80.425~{\rm GeV},~~~~~~~~~m_Z~=~91.1876~{\rm GeV}.
\end{eqnarray}
Here we use the effective values of the light quark masses ($m_u$
and $m_d$) which can reproduce the hardron contribution to the
shift in the fine structure constant $\alpha_{ew}(m_Z^2)$
\cite{DESY}. If we take the electric charge defined in the Thomson
limit $\alpha_{ew} \simeq 1/137.036$, we have
\begin{eqnarray}
   \delta Z_e = -\frac{1}{2}\delta Z_{AA}-\frac{{\rm sin} \theta_W}{2~{\rm cos} \theta_W}\delta
   Z_{ZA},
\end{eqnarray}
and get large radiative corrections for processes at the GeV or
TeV energy scale. In our calculation we use an improved scheme to
make the perturbative calculation more reliable. That means we use
the effective $\overline{MS}$ fine structure constant value at
$Q=m_Z$ as input parameter,
$\alpha_{ew}(m_Z^2)^{-1}|_{\overline{MS}}=127.918$ \cite{RevD}.
This results in the counter-term of the electric charge expressed
as\cite{count1, count2, eberl}
\begin{eqnarray}
\label{ecount}
   \delta Z_e &=&
   \frac{e^2}{6(4\pi)^2}
    \left\{ 4 \sum_f N_C^f e_f^2\left( \Delta+\log\frac{Q^2}{x_f^2} \right)+\sum_{\tilde{f}} \sum_{k=1}^2 N_C^f
    e_{f}^2 \left( \Delta+\log\frac{Q^2}{m^2_{\tilde{f}_k}} \right)     \right.       \nonumber  \\
  &&\left. + 4 \sum_{k=1}^2\left(\Delta+\log\frac{Q^2}{m^2_{\tilde{\chi}_k}}\right)
       +\sum_{k=1}^2\left( \Delta+\log\frac{Q^2}{m^2_{H_k^+}} \right)   \right.      \nonumber   \\
  &&\left.  - 22 \left(\Delta+\log\frac{Q^2}{m_W^2}\right)
  \right\},
\end{eqnarray}
where we take $x_f=m_Z$ when $m_f<m_Z$, and $x_t=m_t$. $e_f$ is
the electric charge of (s)fermion and
$\Delta=2/\epsilon-\gamma+\log4\pi$. $N_C^f$ is color factor and
$N_C^f=1,3$ for (s)leptons and (s)quarks, respectively. The MSSM
parameters are determined by FormCalc package \cite{FormCalc} with
following input parameters:

\par
(i) For the MSSM Higgs sector, we take the CP-odd Higgs boson mass
$M_{A^0}$ and $\tan\beta$ as the input parameters with the
constraint $\tan\beta\geq2.5$. The radiative corrections to Higgs
boson masses up to two-loop contributions have been
involved\cite{Hein}, and we take them as physical masses. The
tree-level Higgs masses can be obtained by using the equations
\begin{eqnarray}
m_{h^0,H^0}^2=\frac{1}{2}\left( M_{A^0}^2+ m_{Z^0}^2\mp
\sqrt{(M_{A^0}^2+ m_{Z^0}^2)^2-4 M_{A^0}^2
m_{Z^0}^2\cos^2(2\beta)} \right), \nonumber \\
m_{H^{\pm}}^2=m_{W}^2+M_{A^0}^2.~~~~~~~~~~~~~~~~~~~
~~~~~~~~~~~~~~~~~~~
\end{eqnarray}
Normally it is necessary to use tree-level Higgs masses through
out the loop calculation to keep the gauge invariance, while for
the phase space integration, the matrix element needs to be
expressed in terms of physical masses for the external
final-states. The way in Ref. \cite{Freit} can handle this
problem. For the specific process $e^+e^- \to t\bar{t}h^0$ in the
MSSM, there is no diagram with exchanging Higgs boson $h^0$ at
tree-level(see Fig.1), and its amplitude does not contain Higgs
mass $m_{h^0}$. Therefore, we need only use tree-level Higgs
masses in the loop integral calculation, and keep the physical
mass $m_{h^0}^{phys}$ in the phase space integration.

\par
(ii) For the sfermion sector, we assume the input parameters as
$M_{\tilde{Q}}=M_{\tilde{U}}=M_{\tilde{D}}=M_{\tilde{E}}=M_{\tilde{L}}=M_{SUSY}$
and the soft trilinear couplings for sfermions $A_q=A_l=A_f$.

\par
(iii) For the chargino and neutralino sector, we take the $SU(2)$
soft-SUSY-breaking gaugino mass parameter $M_2$ and the
Higgsino-mass parameter $\mu$ as the input parameters, and the
$U(1)$ soft-breaking gaugino mass parameter $M_1$ is determined by
adopting the grand unification theory (GUT) relation
$M_1=(5/3)\tan^2\theta_W M_2$ for simplification\cite{Gunion}.

\par
Besides the SM and MSSM input parameters mentioned above, some
more input parameters should be provided in the numerical
calculation, such as the colliding c.m.s. energy $\sqrt{s}$, the
IR regularization parameter $m_\gamma$ and the soft cutoff $\Delta
E/E_b$. In our following calculation, we set the photon mass
regulator $m_\gamma=10^{-2}~{\rm GeV}$ and $\Delta E/E_b=10^{-4}$,
if there is no other statement. In order to show that the full
${\cal O}(\alpha_{ew})$ electroweak relative correction
$\delta_{total}$ is independent of the soft cutoff $\Delta E/E_b$,
we present the relative corrections for the process $e^{+} e^{-}
\to t\bar{t}h^{0}$ as the functions of the soft cutoff $\Delta
E/E_b$ in Fig.3, with $\sqrt{s}=800~{\rm GeV}$, $M_{A^0}=300~{\rm
GeV}$, $\tan\beta=40$, $M_{SUSY}=300~{\rm GeV}$, $M_2=200~{\rm
GeV}$, $\mu=200~{\rm GeV}$ and $A_f=200~{\rm GeV}$. As shown in
the figure, both $\delta_{soft+virtual}$ and $\delta_{hard}$
depend on the soft cutoff $\Delta E/E_b$ obviously, but the full
${\cal O}(\alpha_{ew})$ electroweak relative correction
$\delta_{total}$ is independent of the soft cutoff value. We have
also checked the $m_{\gamma}$ independence numerically.

\begin{figure}[htb]
\centering
\includegraphics[scale=0.5]{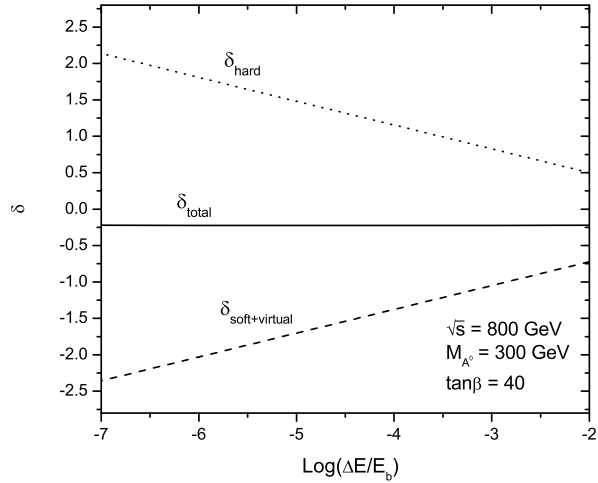}
\caption{The ${\cal O}(\alpha_{ew})$ relative corrections to the
process $e^+ e^- \to t\bar{t}h^0$ as the functions of the soft
cutoff $\Delta E/E_b$.}
\end{figure}

\begin{figure}[htb]
\centering
\includegraphics[scale=0.4]{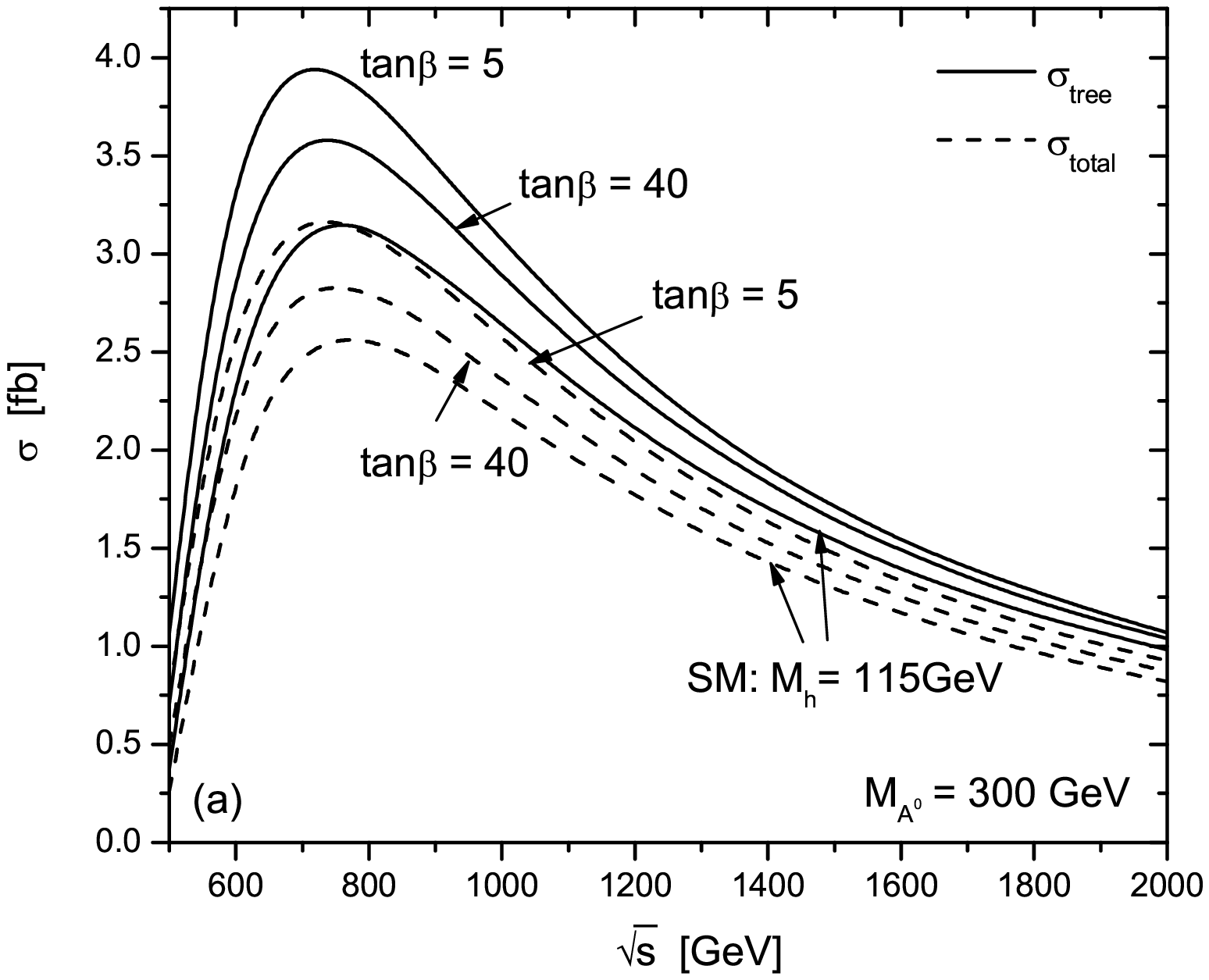}
\includegraphics[scale=0.4]{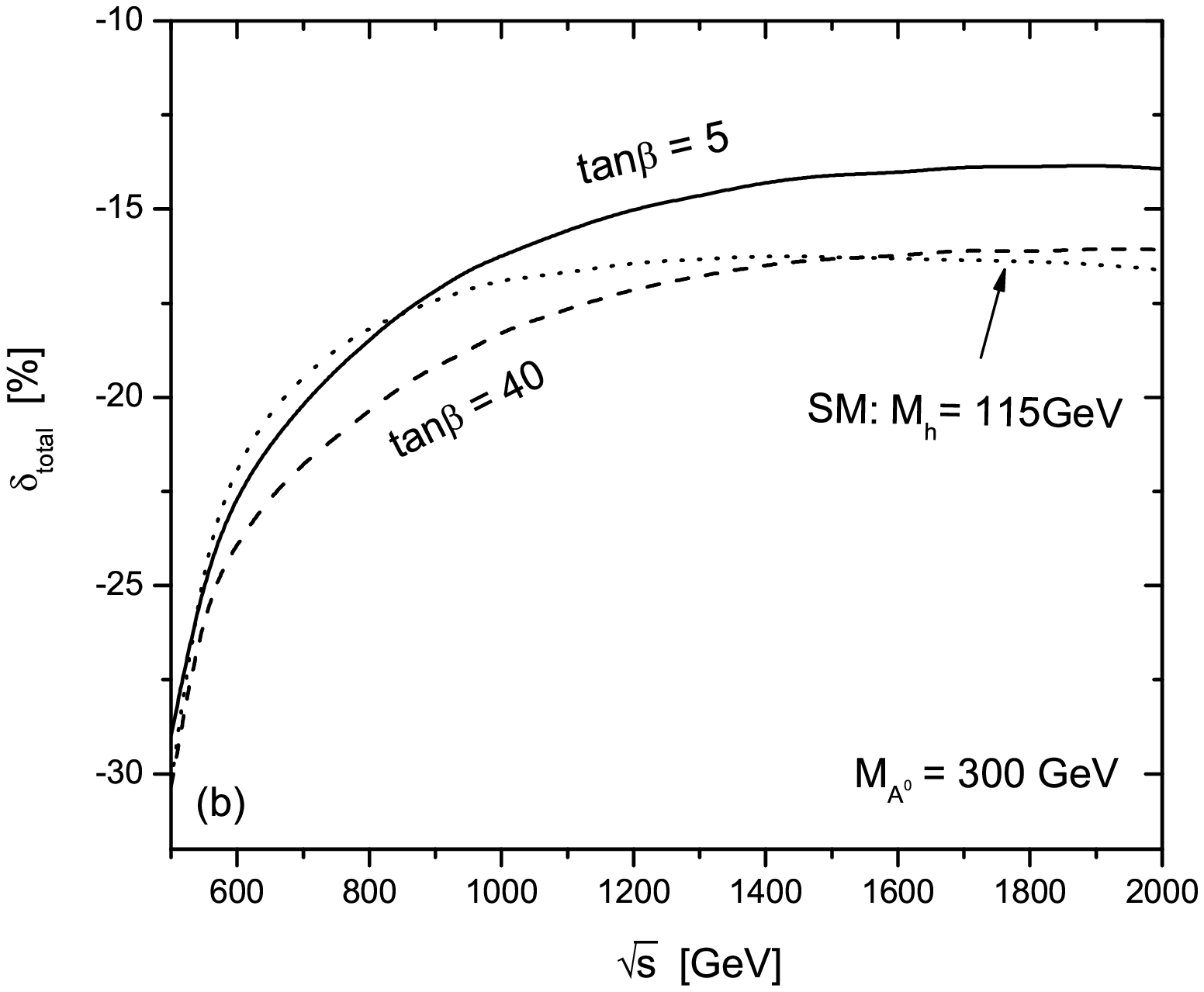}
\caption{The Born and the full one-loop level electroweak
corrected cross sections(shown in Fig.4(a)) as well as the
corresponding relative corrections $\delta_{total}$(shown in
Fig.4(b)) for the process $e^+ e^- \to t\bar{t}h^0$ as the
functions of the c.m.s energy $\sqrt{s}$.}
\end{figure}

\par
By taking the MSSM parameters as $M_{SUSY}=300~{\rm GeV}$,
$M_2=200~{\rm GeV}$, $\mu=200~{\rm GeV}$, $A_f=200~{\rm GeV}$ and
$M_{A^0}=300~{\rm GeV}$, we present Fig.4(a) to show the Born
cross section $\sigma_{tree}$ and the full ${\cal O}(\alpha_{ew})$
corrected cross section $\sigma_{total}$ as the functions of the
c.m.s. energy $\sqrt{s}$ in the SM with $m_h=115~{\rm GeV}$, and
in the MSSM with $\tan\beta=5$ and $\tan\beta=40$, which
correspond to $m_{h^0}=98.36~{\rm GeV}$ for $\tan\beta=5$ and
$m_{h^0}=105.87~{\rm GeV}$ for $\tan\beta=40$, respectively. We
also take the same input parameters as in Refs.\cite{You,Denner},
and get the coincident results for the SM with the corresponding
ones in these references. That comparison is a check for the
correctness of our calculation. In the figure the c.m.s. energy
$\sqrt{s}$ varies from $500~{\rm GeV}$ to $2000~{\rm GeV}$. It
shows that each curve has a peak in the region around the c.m.s.
colliding energy $\sqrt{s} \sim 700~{\rm GeV}$ due to the phase
space feature, and all the curves decrease gently after reaching
their maximal values. We can read out from the figure that the
$\sigma_{tree}$ can reach their own maximum values of $3.96~{\rm
fb}$ and $3.59~{\rm fb}$ at $\sqrt{s} \sim 700~{\rm GeV}$ for
$\tan\beta=5$ and $\tan\beta=40$, respectively, but their maximum
values are shifted to $3.17~{\rm fb }$ and $2.84~{\rm fb}$ after
including the supersymmetric electroweak radiative corrections.
Fig.4(b) shows the dependence of the full ${\cal O}(\alpha_{ew})$
relative correction $\delta_{total}$ on the c.m.s energy
$\sqrt{s}$. There the relative correction increases rapidly with
the increment of the c.m.s. energy in the vicinity of the
threshold energy, but is insensitive to c.m.s. colliding energy
when $\sqrt{s}\gtrsim 1200~{\rm GeV}$.  We present some exact
numerical results of $\sigma_{tree}$, $\sigma_{total}$ and
$\delta_{total}$ in Table 1 by taking above input parameters.

\begin{table}[htb]
\centering
\begin{tabular}{c c c c c c}
\hline $\sqrt{{s}}~[{\rm GeV}]$ & $tan\beta$ & $M_{h^0}[{\rm
GeV}]$ & $\sigma_{tree}[{\rm fb}]$ &
$\sigma_{total}[{\rm fb}]$ & $\delta_{total}[\%]$ \\
\hline
500  &  5  & 98.36  & 1.070746(1) & 0.761(1) & -28.97(9) \\
     &  40 & 105.87 & 0.7086974(7)& 0.4938(6) & -30.33(8) \\
\hline
800  &  5  & 98.36  & 3.808457(3) & 3.105(5) & -18.5(1) \\
     &  40 & 105.87 & 3.515246(3) & 2.800(4) & -20.4(1) \\
\hline
1000 &  5  & 98.36  & 3.065664(3) & 2.568(4) & -16.2(1) \\
     &  40 & 105.87 & 2.889250(3) & 2.362(4) & -18.2(1) \\
\hline
2000 &  5  & 98.36  & 1.073347(1) & 0.924(2) & -13.9(2) \\
     &  40 & 105.87 & 1.041033(1) & 0.874(2) & -16.1(2) \\
\hline
\end{tabular}

\caption{Taking $M_{A^0}=300~GeV$, the Born cross section
$\sigma_{tree}$ and the corrected cross section $\sigma_{total}$
as well as the corresponding relative corrections $\delta_{total}$
for different values of $\tan\beta$ and c.m.s. energy $\sqrt{s}$.}
\end{table}

\begin{figure}[htb]
\vspace*{0cm}
\includegraphics[scale=0.4]{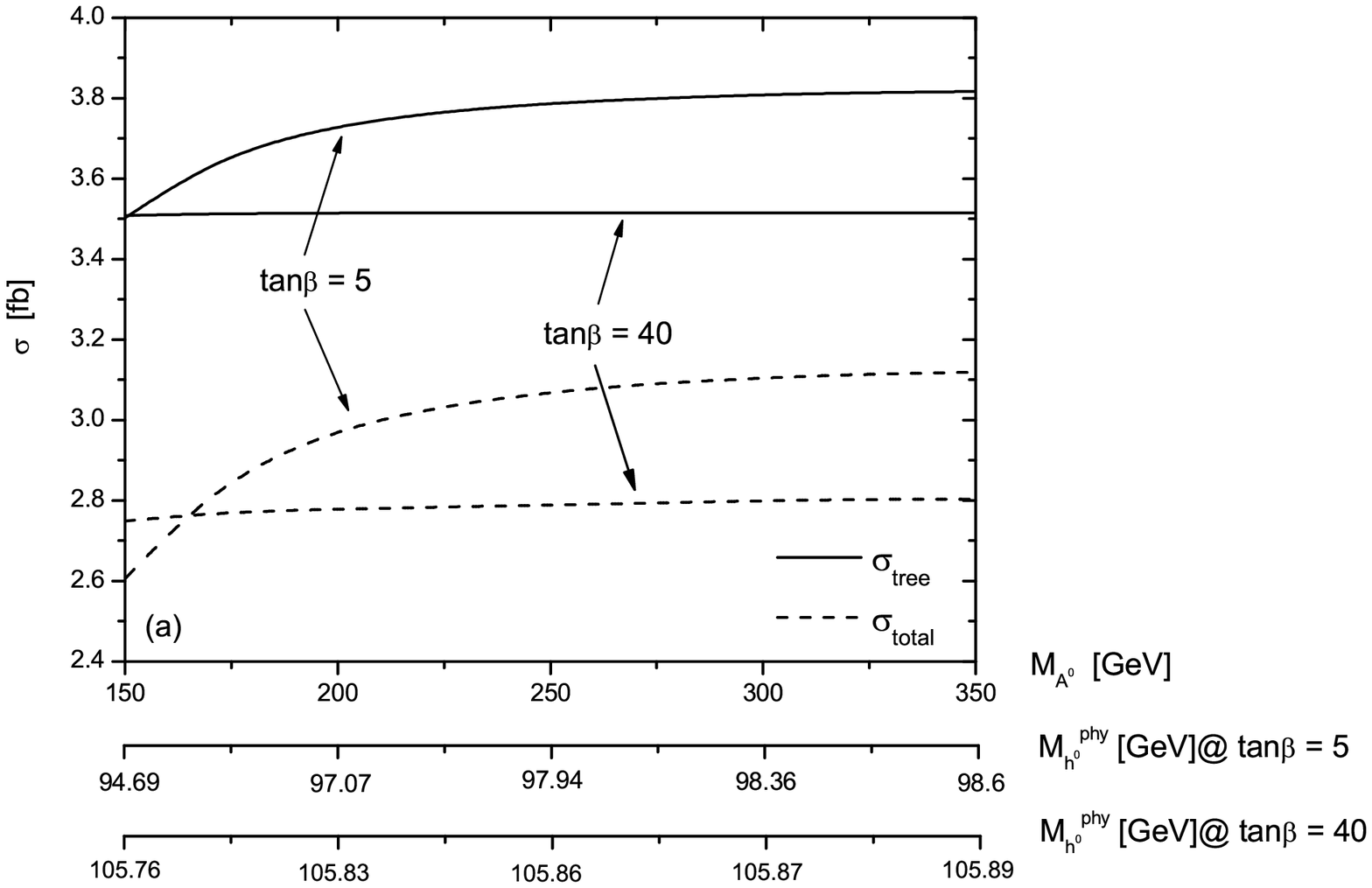}
\includegraphics[scale=0.4]{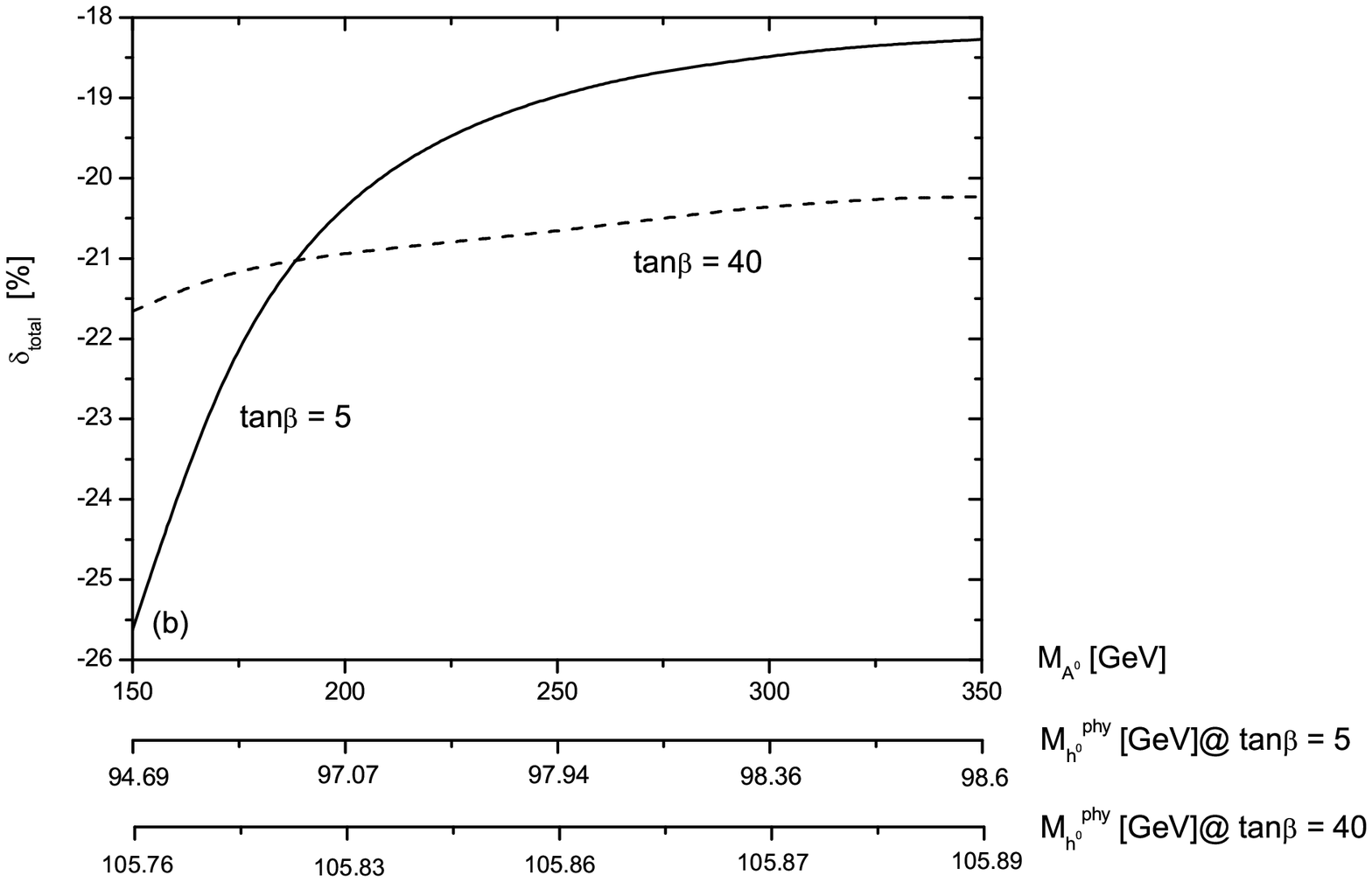}
\vspace*{0cm} \caption{The cross sections at the Born and the
one-loop electroweak levels, and their corresponding relative
corrections for the process $e^+ e^- \to t\bar{t}h^0$ as the
functions of $M_{A^0}(m_{h^0})$ with $\sqrt{s}=800~GeV$ and
$\tan\beta=5,~40$, are shown in Fig.5(a) and Fig.5(b),
respectively.}
\end{figure}

\begin{figure}[htb]
\vspace*{0cm}
\includegraphics[scale=0.4]{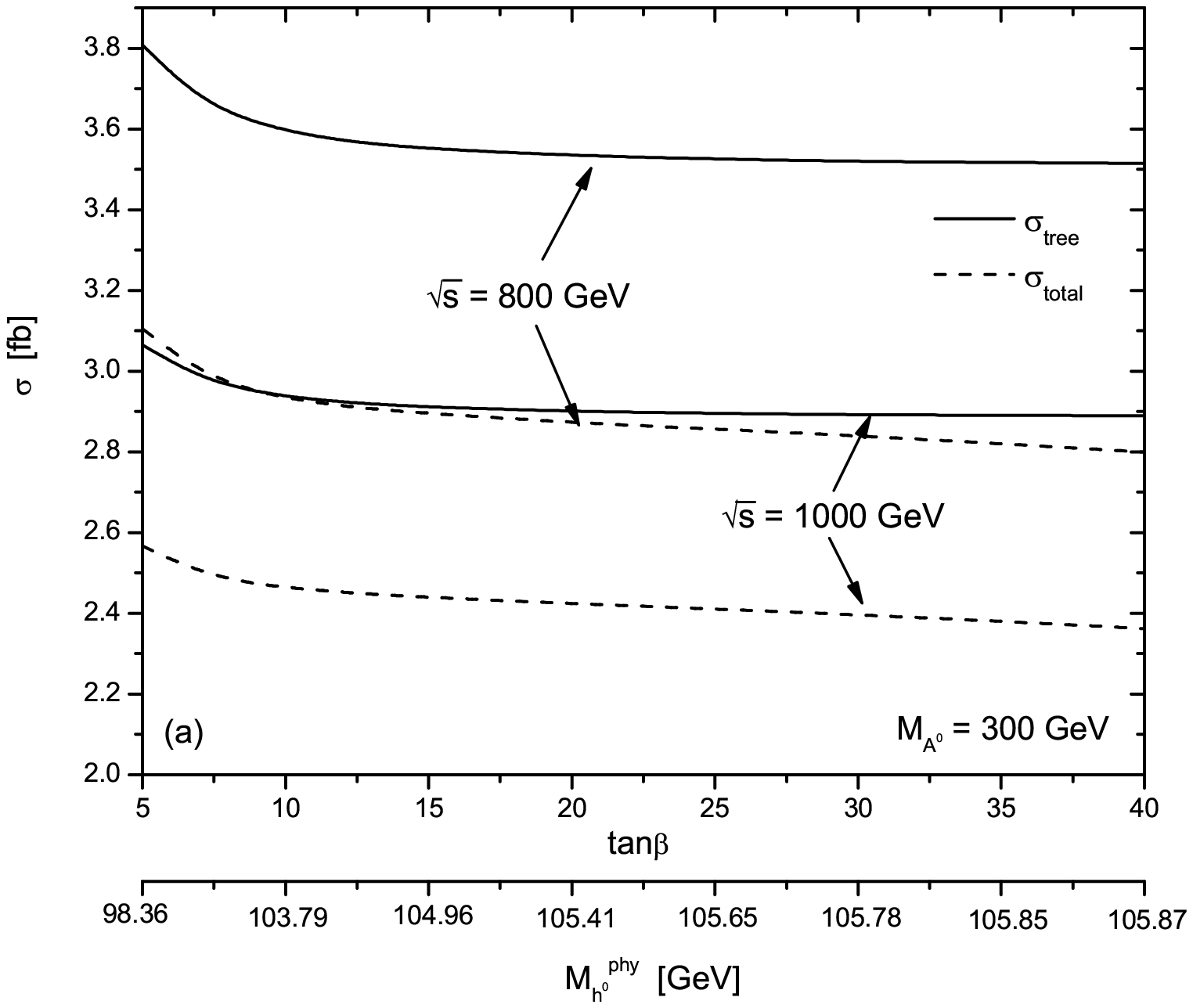}
\includegraphics[scale=0.4]{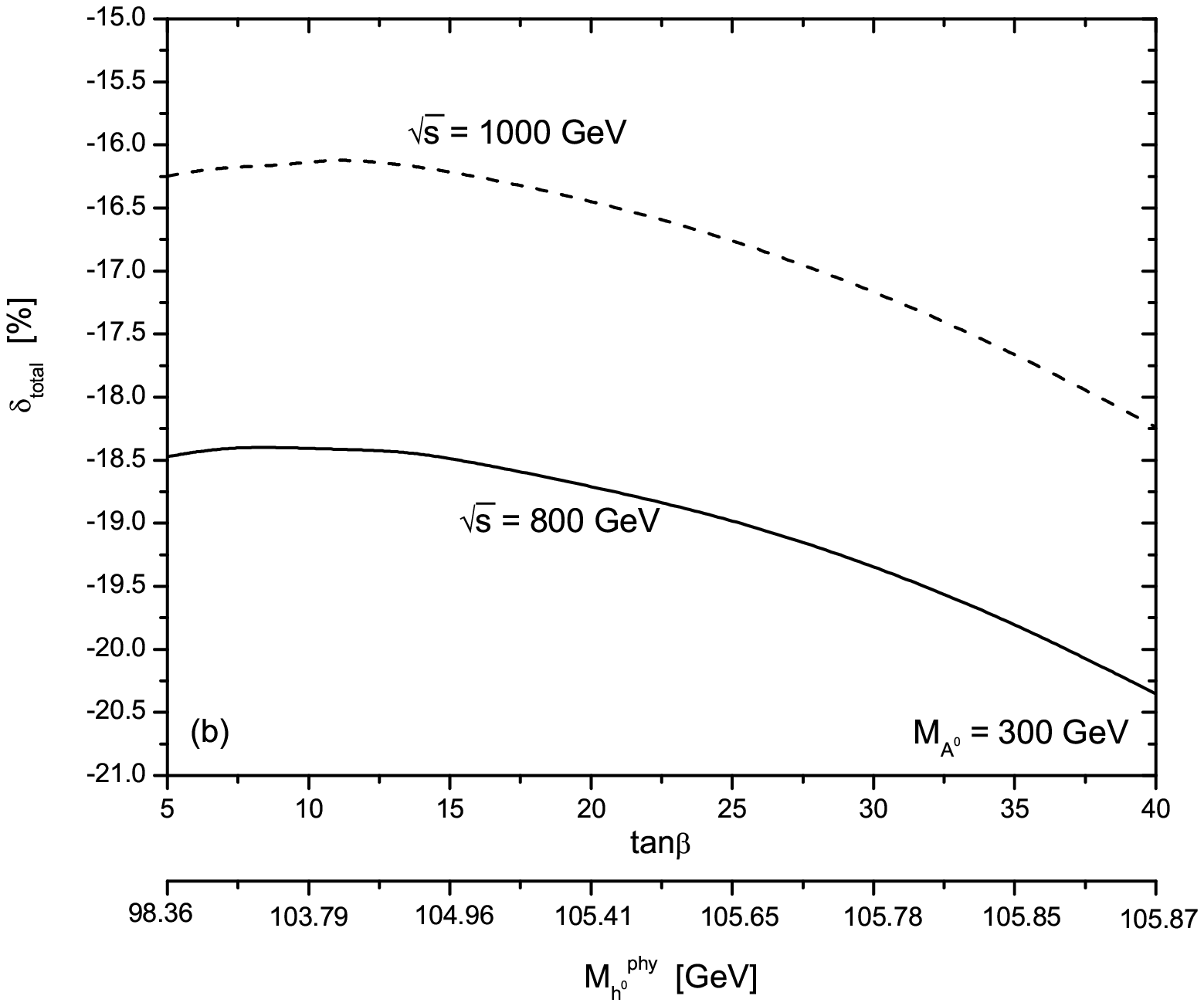}
\caption{The Born and the one-loop level electroweak corrected
cross sections(shown in Fig.6(a)) as well as the corresponding
relative corrections(shown in Fig.6(b)) for the process $e^+ e^-
\to t\bar{t}h^0$ as the functions of $\tan\beta$ with
$\sqrt{s}=800~GeV$ and $\sqrt{s}=1000~GeV$ separately.}
\end{figure}

\par
In Fig.5(a) we present the Born cross section $\sigma_{tree}$ and
the full one-loop electroweak corrected cross section
$\sigma_{total}$ for the process $e^+e^- \to t\bar{t}h^0$ as the
functions of the mass of the CP-odd Higgs boson $A^0$(or
$m_{h^0}$) on the conditions of $M_{SUSY}=300~{\rm GeV}$,
$M_2=200~{\rm GeV}$, $\mu=200~{\rm GeV}$, $\sqrt{s}=800~GeV$ and
$A_f=200~{\rm GeV}$, for $\tan\beta=5$ and $\tan\beta=40$
respectively. The corresponding relative corrections are depicted
in Fig.5(b). As shown in these two figures all the curves of
$\sigma_{tree}$, $\sigma_{total}$ and relative correction
$\delta$, for both $\tan\beta=5$ and $\tan\beta=40$, are less
sensitive to $M_{A^0}(m_{h^0})$, except the relative correction
for $\tan\beta=5$ in the region of $M_{A^0}<250~GeV$. The behavior
for that is due to the fact that when $M_{A^0}$ goes from
$150~GeV$ to $350~GeV$, the phase space of this process does not
change significantly(especially for $\tan\beta=40$), since the
physical mass of $h^0$ varies in a small range from
$94.69~GeV(105.76~GeV)$ to $98.6~GeV(105.89~GeV)$ for
$\tan\beta=5(\tan\beta=40)$ as shown in Fig.5(a) and Fig.5(b).

\par
The Born cross section and the one-loop electroweak corrected
cross section as the functions of $\tan\beta$ are depicted in
Fig.6(a) on the conditions of $M_{A^0}=300~{\rm GeV}$,
$M_{SUSY}=300~{\rm GeV}$, $M_2=200~{\rm GeV}$, $\mu=200~{\rm GeV}$
and $A_f=200~{\rm GeV}$. In this figure both $\sigma_{tree}$
curves for $\sqrt{s}=800~{\rm GeV}$ and $\sqrt{s}=1000~{\rm GeV}$
decrease slowly with the increment of $\tan\beta$ except in the
region of $\tan\beta<10$. To clarify the dependence of the
electroweak relative correction corresponding to Fig.6(a) on
$\tan\beta$, we plot the relative correction versus $\tan\beta$ in
Fig.6(b). One can read from Fig.6(b) that the relative corrections
are again negative as shown in Fig.5, and decrease obviously as
$\tan\beta$ increasing from $10$ to $40$. The values vary from
about $-18.4\%$ to $-20.4\%$ for $\sqrt{s}=800~{\rm GeV}$, and
from $-16.2\%$ to $-18.2\%$ for $\sqrt{s}=1000~{\rm GeV}$ as
$\tan\beta$ running from $5$ to $40$.

\begin{figure}[htbp]
\centering
\includegraphics[scale=0.4]{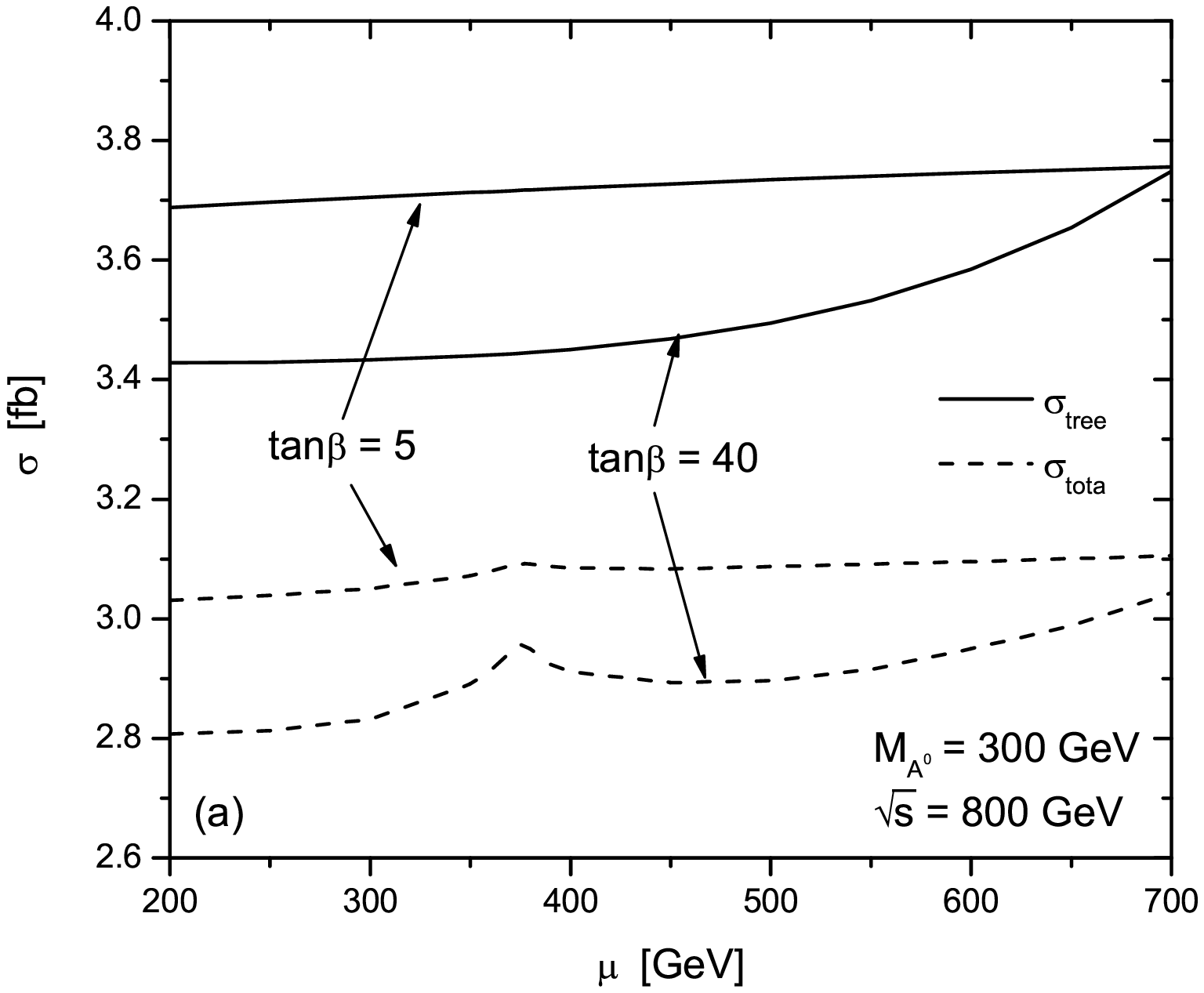}
\includegraphics[scale=0.4]{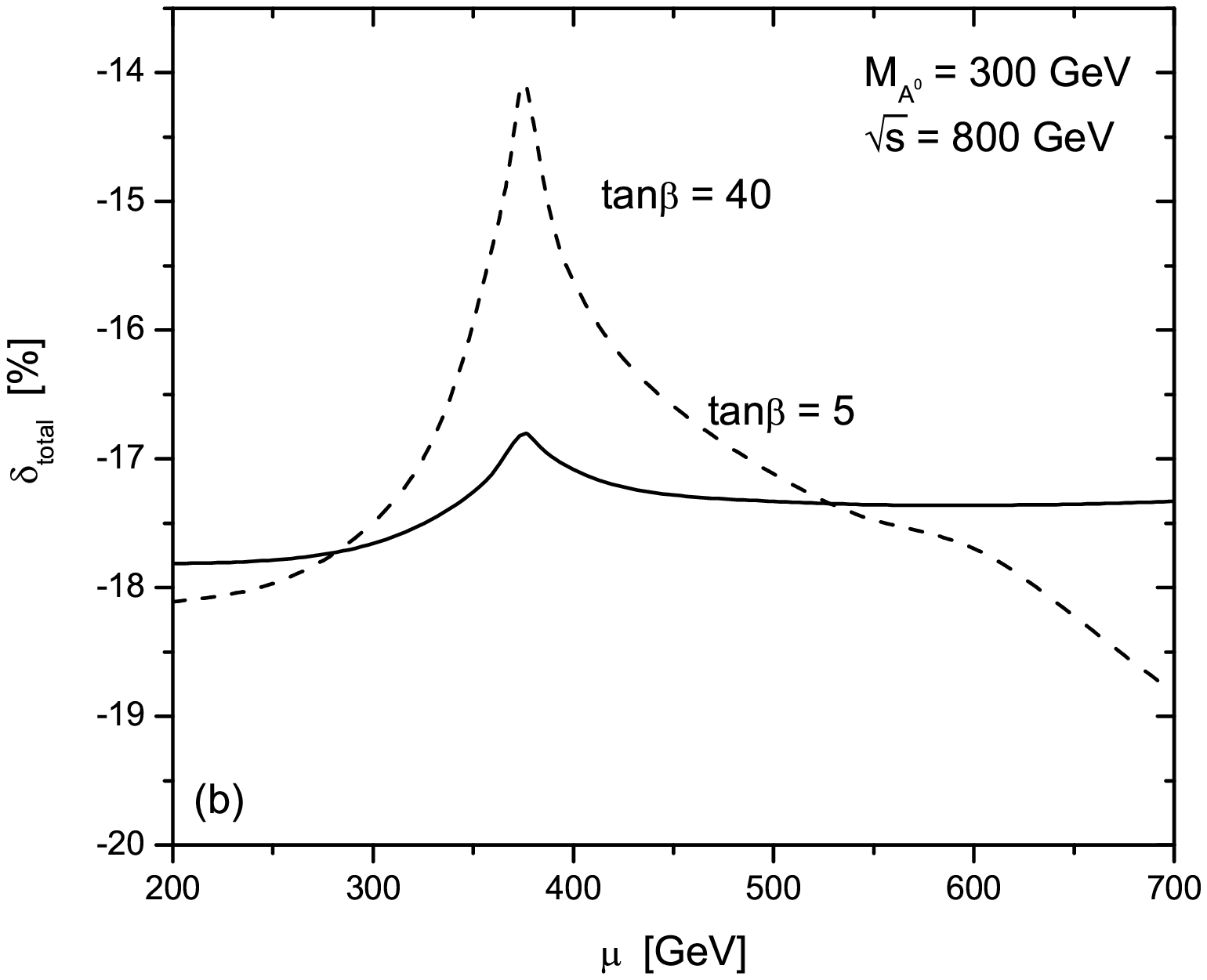}
\caption{The Born and the one-loop level electroweak corrected
cross sections(shown in Fig.7(a)) as well as the corresponding
relative corrections(shown in Fig.7(b)) for the process $e^+ e^-
\to t\bar{t}h^0$ as the functions of the $\mu$ by taking
$\tan\beta =5$ and $\tan\beta=40$ separately.}
\end{figure}

\begin{figure}[htbp]
\centering
\includegraphics[scale=0.4]{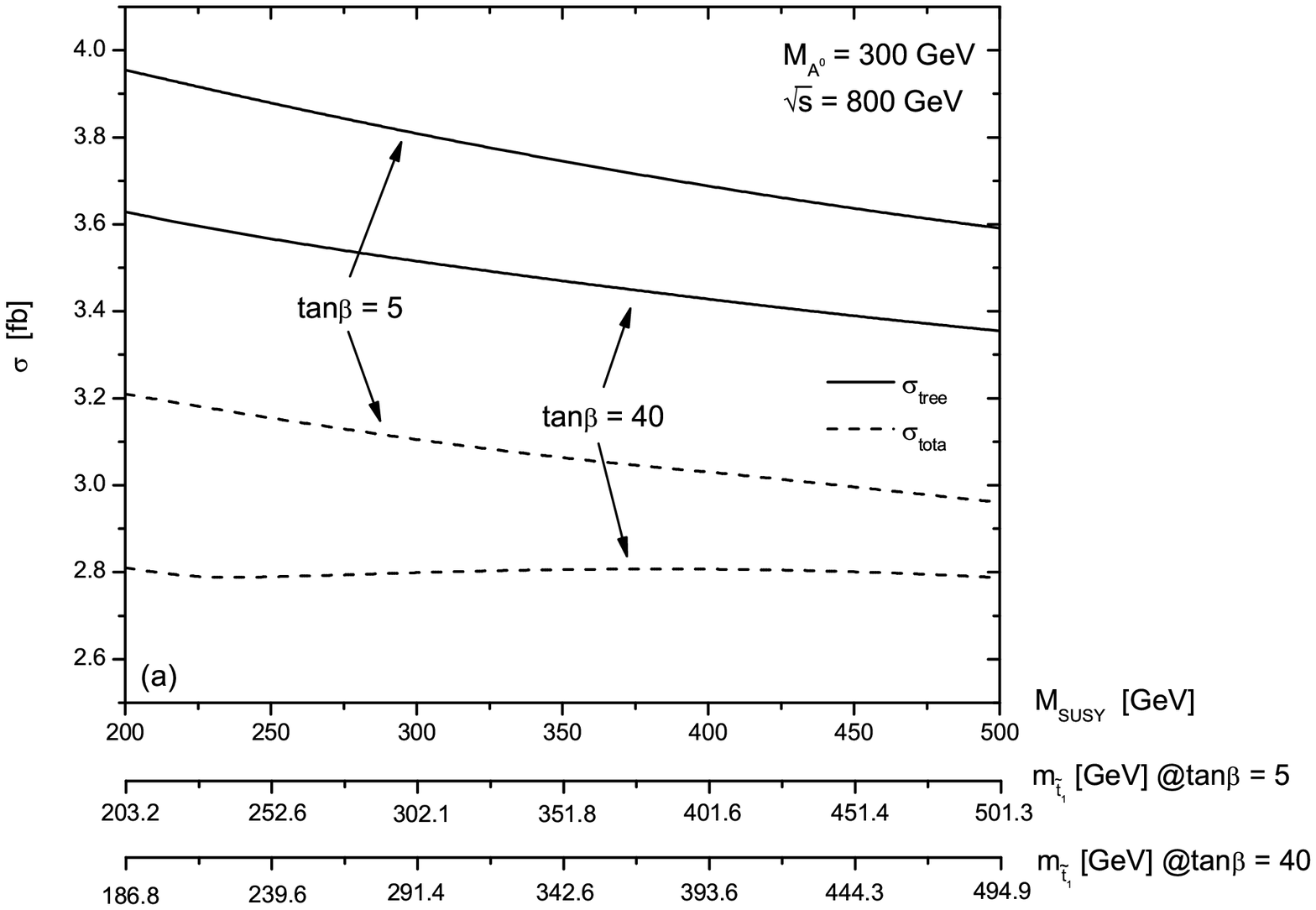}
\includegraphics[scale=0.4]{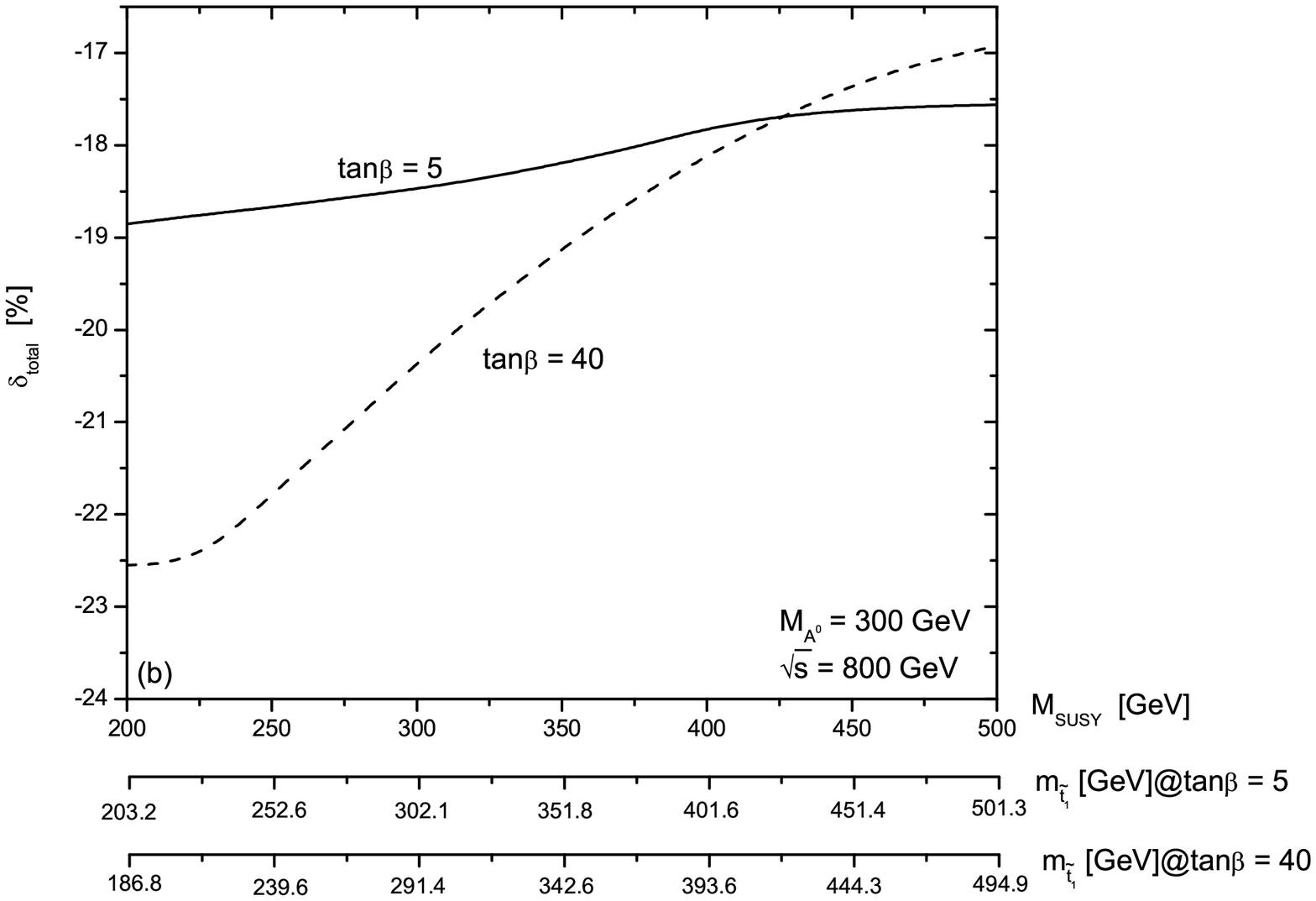}
\includegraphics[scale=0.4]{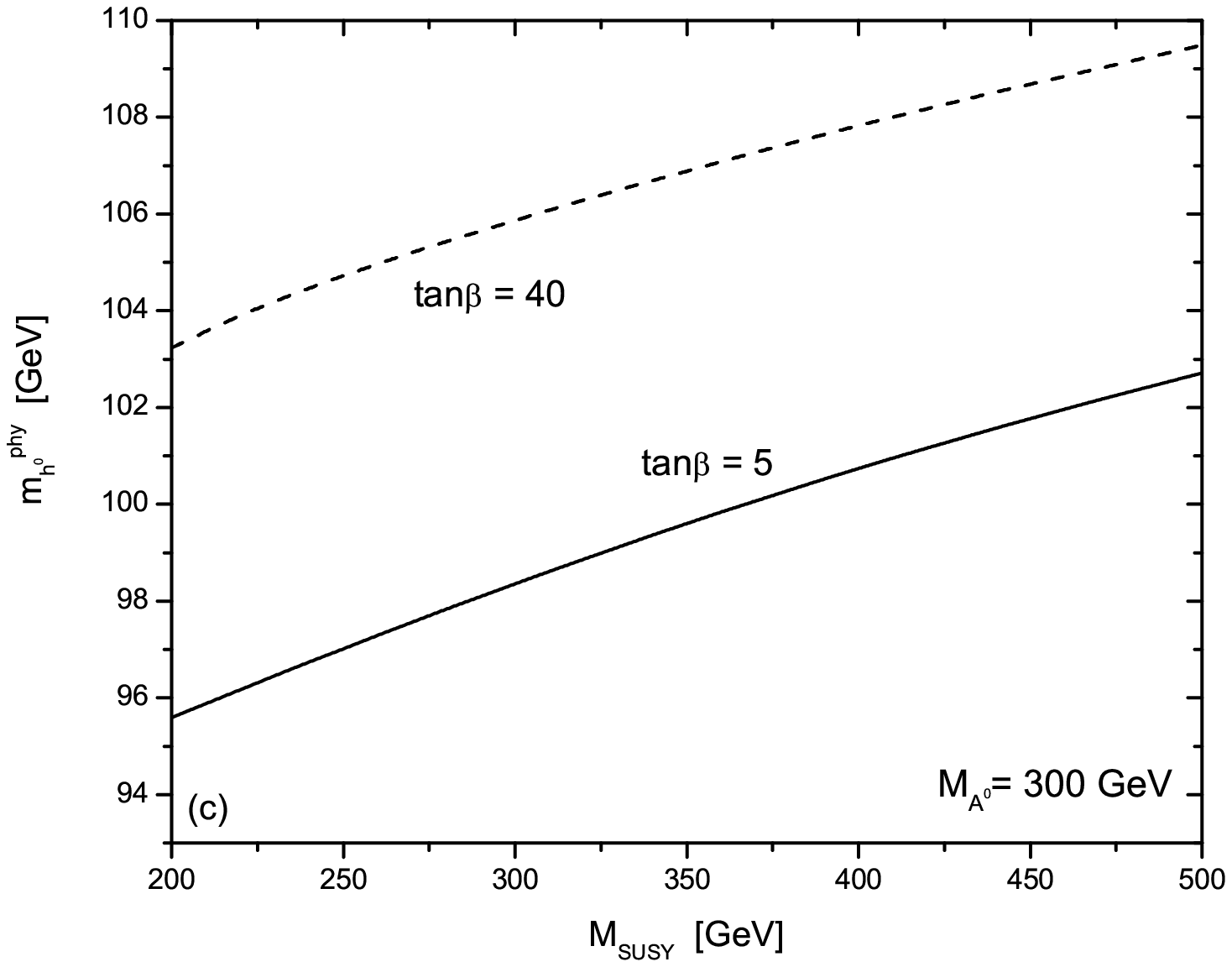}
\caption{The Born and the one-loop level electroweak corrected
cross sections(shown in Fig.8(a)) as well as the corresponding
relative corrections(shown in Fig.8(b)) for the process $e^+ e^-
\to t\bar{t}h^0$ as the functions of the
$M_{SUSY}(m_{\tilde{t}_1})$ by taking $\tan\beta =5$ and
$\tan\beta=40$ separately. The Higgs mass $m_{h^0}$ involving up
to two-loop level radiative corrections as the functions of
$M_{SUSY}$ are plotted in Fig.8(c).}
\end{figure}

\par
In Fig.7(a) we present the Born cross section $\sigma_{tree}$ and
the full ${\cal O}(\alpha_{ew})$ electroweak corrected cross
section $\sigma_{total}$ as the functions of Higgsino-mass
parameter $\mu$ with the conditions of $\sqrt{s}=800~{\rm GeV}$,
$M_{A^0}=300~{\rm GeV}$, $M_{SUSY}=400~{\rm GeV}$, $M_2=200~{\rm
GeV}$ and $A_f=200~{\rm GeV}$. The two full-line curves and two
dashed-line curves in the figure are corresponding to
$\tan\beta=5$ and $\tan\beta=40$ separately. As shown in Fig.7(a),
each curve of $\sigma_{total}$ has a small spike which shows the
resonance effect in the vicinity of $\sqrt{s}\simeq 2
m_{\tilde{\chi}_2^+}$. For $\tan\beta=5$, both $\sigma_{tree}$ and
$\sigma_{total}$ are less sensitive to $\mu$, while for
$\tan\beta=40$, the Born and the electroweak corrected cross
sections increase smoothly with the increment of $\mu$ except in
the range around the resonance peak on the dashed curve for
$\sigma_{total}$. With the same parameter conditions, the
dependence of relative correction on $\mu$ is displayed in
Fig.7(b). For $\tan\beta=5$, the relative correction is also less
sensitive to $\mu$ except in the vicinity of $\mu \sim 377.3~{\rm
GeV}$ for the resonance effect. But the $\delta_{total}$ curve for
$\tan\beta=40$ has a more obvious resonance peak in the vicinity
of $\mu \sim 377.3~{\rm GeV}$, and after arriving the peak value
it decreases rapidly with the increment of $\mu$.
\par
We present the dependence of the Born cross section and the
corrected cross section on the sfermion sector parameter
$M_{SUSY}$(or $ m_{\tilde{t}_1}$) in Fig.8(a), on the conditions
of $\sqrt{s}=800~{\rm GeV}$, $M_{A^0}=300~{\rm GeV}$,
$\mu=200~{\rm GeV}$, $M_2=200~{\rm GeV}$ and $A_f=200~{\rm GeV}$,
with $\tan\beta=5$ and $\tan\beta=40$ respectively. From this
figure we find that both Born cross sections and the electroweak
corrected cross sections decrease slowly with the increment of
$M_{SUSY}$ in the range of $200~{\rm GeV}<M_{SUSY}<500~{\rm GeV}$,
since we take the radiative corrected Higgs mass $m_{h^0}$
involving two-loop corrections as its physical mass. The relations
between the physical Higgs mass $m_{h^0}$ and the soft-SUSY-
breaking mass parameter $M_{SUSY}$ are depicted in Fig.8(c). We
can see from Fig.8(a) and Fig.8(c) that the dependence of the Born
cross-section on $M_{SUSY}$ is due to the Higgs boson mass
$m_{h^0}$ being related to $M_{SUSY}$ at loop level. The relative
corrections as the functions of $M_{SUSY}$ corresponding to
Fig.8(a) are depicted in Fig.8(b). In contrast to the case of
$\tan\beta=5$, the full ${\cal O}(\alpha_{ew})$ electroweak
relative correction for $\tan\beta=40$ is more sensitive to
parameter $M_{SUSY}$. The electroweak relative correction for
$\tan\beta=40$ varies in the range between $-22.6\%$ and $-16.9\%$
when $M_{SUSY}$ goes from $200~GeV$ to $500~GeV$.

\begin{figure}[htb]
\centering
\includegraphics[scale=0.35]{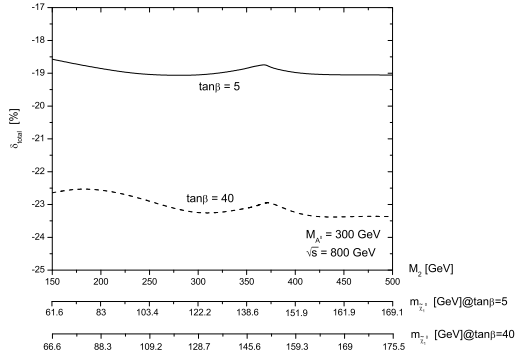}
\vspace*{0cm} \caption{The electroweak relative corrections for
the process $e^+ e^- \to t\bar{t}h^0$ as the functions of the
$M_2$ with $\tan\beta =5$ and $\tan\beta=40$, respectively.}
\end{figure}

\begin{figure}[htb]
\centering
\vspace*{-1cm}
\includegraphics[scale=0.4]{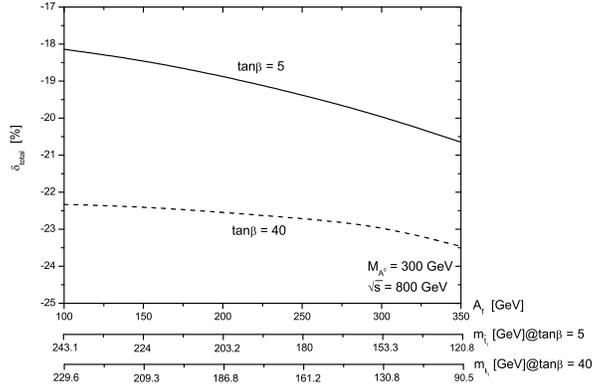}
\vspace*{0cm} \caption{The electroweak relative corrections for
the process $e^+ e^- \to t\bar{t}h^0$ as the functions of the
$A_f$ with $\tan\beta =5$ and $\tan\beta=40$, respectively.}
\end{figure}

\par
In Fig.9 and Fig.10, we depict the dependence of the full ${\cal
O}(\alpha_{ew})$ electroweak relative correction on the gaugino
mass parameter $M_2$ (or neutralino mass $m_{\tilde{\chi}^0_1}$)
and the soft trilinear couplings for sfermions $A_f$ (or scalar
top-quark mass $m_{\tilde{t}_1}$) respectively. There we take the
input parameters as $\sqrt{s}=800~{\rm GeV}$, $M_{A^0}=300~{\rm
GeV}$, $\mu=200~{\rm GeV}$, $M_{SUSY}=200~{\rm GeV}$ and
$A_f=200~{\rm GeV}$ for Fig.9 and $\sqrt{s}=800~{\rm GeV}$,
$M_{A^0}=300~{\rm GeV}$, $\mu=200~{\rm GeV}$, $M_{SUSY}=200~{\rm
GeV}$ and $M_2=200~{\rm GeV}$ for Fig.10. In Fig.9, each curve has
a small peak at about $M_2 \sim 370~GeV$. That reflects the
resonance effect satisfying the condition of $\sqrt{s} \sim 2
m_{\tilde{\chi}_2^+}$. From these two figures we can see that the
one-loop electroweak relative correction are less sensitive to
$M_2$(or $m_{\tilde{\chi}^0_1}$) and $A_f$(or $m_{\tilde{t}_1}$)
quantitatively. The variation ranges of the relative corrections
for both curves in Fig.9 are less than $1\%$ in our chosen
parameter space. Fig.10 shows that the variations of the relative
corrections for curves of $\tan\beta=40$ and $\tan\beta=5$ are
less than $1\%$ and $3\%$ respectively, when $A_f$ goes from
$100~GeV$ to $350~GeV$.

\vskip 10mm
\section{Summary}
In this paper, we present the calculation of the full ${\cal
O}(\alpha_{ew})$ electroweak correction to the process $e^+e^-\to
t\bar{t}h^0$ at a LC in the MSSM. We analyze the numerical results
and investigate the dependence of the cross section and relative
correction on $\sqrt{s}$ and several MSSM parameters. We find that
these corrections generally reduce the Born cross sections and the
relative correction is typically of order $-20\%$. The electroweak
relative correction is strongly related to $\tan\beta$, and has
obvious dependence on $M_{A^0}$ and $M_{SUSY}$ on the conditions
of $\tan\beta=5$ and $\tan\beta=40$, respectively. The results
also show that the one-loop electroweak relative correction is
generally less sensitive to $M_{2}$ and $A_{f}$ in the range of
$150~{\rm GeV} <M_{2}<500~{\rm GeV}$ and $100~{\rm GeV}
<A_{f}<350~{\rm GeV}$, respectively. We conclude that the complete
${\cal O}(\alpha_{ew})$ electroweak corrections to the process
$e^+e^-\to t\bar{t}h^0$ are generally significant and cannot be
neglected in the precise experiment analysis.

\vskip 10mm \noindent{\large\bf Acknowledgement:}

This work was supported in part by the National Natural Science
Foundation of China and a special fund sponsored by China Academy
of Science.

\vskip 5mm%\noindent{\large {\bf Appendix}}


\vskip 10mm
\begin{thebibliography}{s25}
\bibitem{sm1}
         S. L. Glashow, Nucl. Phys. {\bf B22}, 579 (1961) ;
         S. Weinberg, Phys. Rev. Lett. {\bf 19}, 1264 (1967);
         A. Salam, in \textit{Proceedings of the 8th Nobel Symposium}, Stockholm, 1968,
         edited by N. Svartholm
         (Almquist and Wiksells, Stockholm, 1968), p.367;
         H. D. Politzer, Phys. Rep. {\bf 14} 129 (1974).

\bibitem{sm2}
         P. W. Higgs, Phys. Lett {\bf 12}, 132 (1964), Phys. Rev. Lett. {\bf 13}, 508 (1964);
         Phys. Rev. {\bf 145}, 1156 (1966);
         F. Englert and R. Brout, Phys. Rev. Lett. {\bf 13}, 321 (1964);
         G. S. Guralnik, C. R. Hagen, and T. W. B. Kibble, \textit{ibid}. {\bf 13}, 585 (1964);
         T. W. B. Kibble, Phys. Rev. {\bf 155}, 1554 (1967).

\bibitem{ALEPH1}
        (ALEPH), (DELPHI), (L3) and (POAL) Collaborations, the LEP working group for
	  Higgs boson searches'LHWG Note 2002-01 (2002),
        in ICHEP'02 {\it Amsterdam}, 2002; and additional updates at
        http://lephiggs.web.cern.ch/LEPHIGGS/www/Welcome.html; P.
        A. McNamara and S. L. Wu, Rept. Prog. Phys. {\bf 65}, 465
        (2002).

\bibitem{LEP}
           The LEP Collaborations: ALEPH Collaboration, DELPHI Collaboration,
           L3 Collaboration, OPAL Collaboration, the LEP Electroweak Working Group,
          the SLD Electroweak, Heavy Flavour Groups, CERN-PH-EP/2004-069,
          LEPEWWG/2004-01, arXiv:hep-ex/0412015.

\bibitem{ALEPH2}
        (ALEPH), (DELPHI), (L3) and (POAL) Collaborations, the LEP working group for
	   Higgs boson searches'LHWG Note 2002-04, LHWG Note
        2002-05.

\bibitem{higgs}
        The LEP Collaborations: ALEPH Collaboration, DELPHI Collaboration, L3 Collaboration,
          OPAL Collaboration, the LEP Electroweak Working Group, the SLD Heavy Flavour Working Group,
          LEPEWWG/2002-02, CERN-EP/2002-091, hep-ex/0212036;
         S. Alekhin et.al., 'The QCD/SM Working Group: Summary Report', arXiv:hep-ph/0204316.

\bibitem{TESLA}
        'TESLA: The superconducting electron positron linear collider with an
        integrated X-ray laser laboratory. Technical design report, Part 2:
        The Accelerator', Report No. DESY-01-11, edited by R. Brinkmann, K. 
        Flottmann, J. Rossbach, P. Schmuser, N. Walker, and H. Weise, 2001 (unpublished).

\bibitem{NLC}
        C. Adolphsen {\it et al.,} (International Study Group Collaboration),
        'International study group progress report on linear collider development',Report Nos.
        SLAC-R-559 and KEK-REPORT-2000-7, 2000 (unpubplished).

\bibitem{JLC}
        N. Akasaka {\it et al.,} 'JLC design study', Report No. KEK-REPORT-97-1.

\bibitem{CLIC}
        'A 3 TeV $e^+e^-$ Linear Collider Based on
         CLIC Technology',Report No. CERN-2000-008, edited by G. Guignard (unpublished).

\bibitem{Reina1}
        L. Reina and S. Dawson, Phys. Rev. Lett. {\bf 87}, 201804
        (2001).

\bibitem{Reina2}
        L. Reina, S. Dawson and D. Wackeroth, Phys. Rev. {\bf D65},
        053017 (2002).

\bibitem{Been}
        W. Beenakker, S. Dittmaier, M. Kr\"{a}mer, B. Plumper,
        M. Spira and P. M. Zerwas, Phys. Rev. Lett. {\bf 87}, 201805
        (2001); Nucl. Phys. {\bf B653}, 151 (2003).

\bibitem{Rainwater}
        D. Rainwater, M. Spira and D. Zeppenfeld, Report No. MAD-PH-02-1260 (unpublished)..

\bibitem{You}
        Y. You, W.-G. Ma, H. Chen, R.-Y. Zhang, Y.-B. Sun, H.-S. Hou, Phys. Lett.
        {\bf B571} (2003) 85, arXiv:hep-ph/0306036; G. Belanger, F. Boudjema, J. Fujimoto,
        T. Ishikawa, T. Kaneko, K. Kato, Y. Shimizu and Y. Yasui, Phys. Lett. {\bf B571} (2003)163,
        arXiv:hep-ph/0307029; A. Denner, S. Dittmaier, M. Roth, M. M. Weber, Phys. Lett.
        {\bf B575} (2003)290, arXiv:hep-ph/0307193.

\bibitem{Denner}
        A. Denner, S. Dittmaier, M. Roth and M. M. Weber, Nucl.Phys. {\bf B680} 85 (2004);
        Eur. Phys. J. {\bf C33} S635 (2004); C. Farrell, A. H.
        Hoang, Phys. Rev. {\bf D72} (2005) 014007, arXiv:hep-ph/0504220.

\bibitem{Wu}
        X. H. Wu, C. S. Li and J. J. Liu, arXiv:hep-ph/0308012.

\bibitem{Zhu}
        S. H. Zhu, arXiv:hep-ph/0212273.

\bibitem{Spira}
        P. H\"{a}fliger and M. Spira, arXiv:hep-ph/0501164.

\bibitem{Cheung}
        K. Cheung, Phys. Rev. {\bf D47} (1993) 3750.

\bibitem{Chen}
        H. Chen, W.-G. Ma, R.-Y. Zhang, P.-J. Zhou, H.-S. Hou and Y.-B. Sun, Nucl. Phys. {\bf B683} 196
        (2004).

\bibitem{loop}
        G. Passarino and M. Veltman, Nucl. Phys. {\bf B160} 151
        (1979).

\bibitem{Feynarts}
        T. Hahn, Comp. Phys. Commun. {\bf 140} 418 (2001).

\bibitem{Form}
        J. A. M. Vermaseren, arXiv:math-ph/0010025.

\bibitem{pentagon}
        A. Denner and S. Dittmaier, Nucl. Phys. {\bf B658} 175
        (2003).

\bibitem{RevD}
        S. Eidelman, \textit{et al.}, Phys. Lett. {\bf B592} 1
        (2004).

\bibitem{ren}
        J. Guasch, W. Hollik, J. Sola, J. High Energy Phys. 10 (2002) 040, arXiv:hep-ph/0207364;
        W. Hollik, E. Kraus, M. Roth, C. Rupp, K. Sibold and D. Stoecklinger,
        Nucl.Phys. B639 (2002) 3-65, arXiv:hep-ph/0204350.

\bibitem{MSSM}
        D. Pierce and A. Papadopoulos, Phys. Rev. {\bf D47} 222
        (1993); R.-Y. Zhang, W.-G. Ma, L.-H. Wan and Y. Jiang, Phys. Rev. {\bf D65}
        (2002)075018.

\bibitem{Harris} B.W. Harris and J.F. Owens, Phys. Rev. D {\bf 65}
        (2002) 094032

\bibitem{Hooft}
        G. 't Hooft and M. Veltman, Nucl. Phys. {\bf B153} 365
        (1979).

\bibitem{grace}
        T. Ishikawa, T. Kaneko, K. Kato, S. Kawabata, Y. Shimizu and H. Tanaka,
        "GRACE manual", KEK report 92-19, 1993(unpublished)..

\bibitem{DESY}
        F. Jegerlehner, Report No. DESY 01-029 (unpublished).

\bibitem{count1}
        C. Weber, H. Eberl, W. Majerotto, Phys. Lett. {\bf B572} 56 (2003).

\bibitem{count2}
        H. Eberl, M. Kincel, W. Majerotto and Y. Yamada, Nucl. Phys. {\bf B625} 372 (2002).

\bibitem{eberl}
        K. Kova\v{r}\'{\i}k, C. Weber, H. Eberl, W. Majerotto, Phys. Lett. {\bf B591} 242 (2004).

\bibitem{FormCalc}
        Thomas Hahn and Christian Schappacher, Comput. Phys. Commun. 143 54 (2002).

\bibitem{Hein}
        S. Heinemeyer, W. Hollik, G. Weiglein, Phys.Lett. B455, 179 (1999).

\bibitem{Freit}
         A. Freitas, A. van Manteuffel and P.M. Zerwas, Eur.Phys.J. C40 435 (2005),          arXiv:hep-ph/0408341.

\bibitem{Gunion}
        J. F. Gunion, H. E. Haber, Nucl. Phys. {\bf B272} 1 (1986) .

\end{thebibliography}
\end{document}